# Preserving Surface Strain in Nanocatalysts via Morphology Control

Chuqiao Shi[1†], Zhihua Cheng[2†], Alberto Leonardi[3,4,5], Yao Yang[6], Michael Engel[4], Matthew R. Jones[1,2#], Yimo Han[1#]

[1] Department of Materials Science and NanoEngineering, Rice University, Houston, TX, 77006, United States

[2] Department of Chemistry, Rice University, Houston, TX, 77006, United States

[3] Diamond Light Source Ltd., Harwell Science and Innovation Campus, Didcot, Oxfordshire, OX11 0DE, United Kingdom.

[4] Institute for Multiscale Simulation, IZNF, Friedrich-Alexander-Universität Erlangen-Nürnberg, 91058 Erlangen, Germany.

[5] Department of Earth and Atmospheric Sciences, Indiana University, 1001 East 10th Street, Bloomington, Indiana, IN47405-1405, United States.

[6] Department of Chemistry and Chemical Biology, Cornell University, Ithaca, NY 14850, United States.

† These authors contributed equally to this work.

[#] To whom correspondence should be addressed:

Yimo Han (Email: yimo.han@rice.edu), Matthew R. Jones (Email: mrj@rice.edu)

**Abstract:**

Engineering strain critically affects the properties of materials and has extensive applications in semiconductors and quantum systems. However, the deployment of strain-engineered nanocatalysts faces challenges, particularly in maintaining highly strained nanocrystals under reaction conditions. Here, we introduce a morphology-dependent effect that stabilizes surface strain even under harsh reaction conditions. Employing four-dimensional scanning transmission electron microscopy (4D-STEM), we discovered that core-shell Au@Pd nanoparticles with sharp-edged morphologies sustain coherent heteroepitaxial interfaces with designated surface strain. This configuration inhibits dislocation due to reduced shear stress at corners, as molecular dynamics simulations indicate. Demonstrated in a Suzuki-type cross-coupling reaction, our approach achieves a fourfold increase in activity over conventional nanocatalysts, owing to the enhanced stability of surface strain. These findings contribute to advancing the development of advanced nanocatalysts and indicate broader applications for strain engineering in various fields.

**Introduction**

Engineering surface strain can improve the performance of nanocatalysts as it directly influences electronic states and active sites, offering the potential to design high-performance catalysts for a wide range of applications(*1–5*). Among the various methods to impart surface strain, the growth of core-shell nanostructures has emerged as a promising approach since it enables precise control of surface strain through lattice heteroepitaxy(*6–9*). However, the propensity of epitaxial strain to spontaneously relax through dislocation nucleation under operating conditions limits their utility. Consequently, the discovery of methods to trap metastable strain states in materials are desirable for improving their durability and performance.

Matthews and Blakeslee's equilibrium theory of dislocation formation(*10*) suggests that strain relaxation in heteroepitaxial growth begins when the film exceeds a certain *critical thickness* that occurs when the accumulated strain energy surpasses that required to create misfit dislocations. For core-shell metal nanoparticles (NPs), previous research has suggested a *critical thickness* of only a few monolayers, owing to the ease of forming misfit dislocations and stacking faults that release the surface strain(*11–14*). Thus, great synthetic effort has been devoted to the generation and maintenance of conformal, ultra-thin heteroepitaxial shells that nonetheless tend to quickly relax due to their thermodynamic instability in-operando, thereby rapidly losing their effectiveness(*15–21*). This is a particularly salient factor in strain-engineered catalytic systems which often require high temperatures and/or harsh reaction conditions to perform optimally.

In this work, we demonstrate sharp-edge morphology in the core effectively inhibits dislocation formation in the shell, thereby improving surface strain stability compared to their conventional rounded-core counterparts (**Fig. 1a**). We use core-shell $Au_{cube}@Pd_{cube}$ particles as a model system, as shown by the energy dispersive X-ray spectroscopy (EDX) maps (**Fig. S1**), given the large but accommodatable lattice mismatch of 4.8% between relaxed Au and Pd lattices(*12*). Atomic-resolution annular dark field STEM (ADF-STEM) images (**Fig. 1b**) depict core-shell $Au_{cube}@Pd_{cube}$ NPs that represent the extremes of radius of curvature values of 4.4 nm and 18.5 nm at their corners. For the sharp-core Au NPs, the Pd shell forms a coherent interface, while those with rounded-cores show mismatch near the corners.

**Results**

In order to understand the differences in strain states in these two cases, we quantify it *via* a high-precision whole-particle strain analysis using 4D-STEM. Different from conventional geometric phase analysis (GPA)(*5*, *22*), which requires atomic-resolution images and typically offers a limited field of view with high noise, we use a nanobeam 4D-STEM approach(*23*, *24*), which affords us a remarkable (< 0.18%) strain precision and allows for comprehensive mapping of entire particles over dimensions spanning from a few to hundreds of nanometers (**Fig. 1c**). Throughout the collection of 4D datasets, the electron probe scans across the core-shell Au$_{cube}$@Pd$_{cube}$ NP, while the electron microscope pixel array detector (EMPAD)(*25*) records full diffraction patterns at each position (**Fig. 1d**, top). Although core-shell systems pose challenges in 4D data processing, including lattice misalignments and projection complexities arising from overlapped materials, we addressed them via optimizing exit-wave power-cepstrum (EWPC) analysis(*26*) (**Fig. 1d**, bottom, with details in **Fig. S2** and **Methods**) and developing mismatch correction for the projected core (details in **Fig. S3** and **Methods**), respectively. These steps are essential in eliminating potential errors from the strain profile, ultimately yielding high-precision strain measurements spanning the entirety of the nanostructure.

The strain map for individual sharp-core Au$_{cube}$@Pd$_{cube}$ particles (**Fig. 1e** and **f**) reveals a highly strained 8-nm Pd shell surrounding the Au core. Since the strain is calculated with respect to a perfect Au crystal as a reference, 0% and -4.8% correspond to the relaxed values of Au and Pd lattices, respectively. The measured lattice constants of Au and Pd along the interface are both close to the ideal Au lattice constant (statistics shown in **Fig. 1g** and calculation in **Fig. S4a-c**). Conversely, the Pd lattice undergoes compressive strain in the direction perpendicular to the interface due to the Poisson effect (as shown in **Fig. S5**). This tetragonal lattice deformation in the Pd shell is best illustrated by calculating the lattice constant ratio $a_x/a_y$ (as shown in **Fig. S6**), where $a_x$ and $a_y$ are the lattice constants along x and y directions, respectively. Despite a similar shell thickness to the sharp-core Au$_{cube}$@Pd$_{cube}$ NPs, those with rounded-cores show notable differences in the strain maps that indicate a release of tensile strain (**Fig. 1h** and **i**), rendering the interface between the Au and Pd incoherent. In this case, the measured Pd shell lattice constant (3.970 Å) is 2.7% mismatched with Au and closer to the bulk value for Pd (statistics shown in **Fig. 1j** and calculation in **Fig. S4d-f**), suggesting strain relaxation. This conclusion is corroborated by maps of the off-diagonal components of the stress-strain tensor which reveal substantial residual tangential shear and rotational effects for sharp-core Au$_{cube}$@Pd$_{cube}$ NPs (**Fig. 1f**), that are not present in the round-core Au$_{cube}$@Pd$_{cube}$ NPs (**Fig. 1i**). This further illustrates the sensitive dependence of strain preservation on particle core morphology in heteroepitaxial nanostructures.

To reveal the role of core sharpness on strain relaxation, we developed synthetic methods that allowed for the growth of Au$_{cube}$@Pd$_{cube}$ NPs with the same Pd shell thickness but different Au core tip curvatures. Briefly, by increasing the bromide ion concentration during particle synthesis, growth along <111> directions can be accelerated, leading to the development of Au cubes with predominantly {100} facets and sharper tips (details in **Fig. S7** and **Methods**). We quantify Au core morphology using the *sharpness index* (*SI*) (*27*) (defined in **Fig. S8**), which compares the tip radius of curvature to the edge length with values closer to one indicating sharper features. To precisely control Pd shell growth, we manipulated the quantity of Au cubes added to a Pd overgrowth solution (see **Methods** for more details). Careful control of these parameters allowed for the measurement of 4D-STEM strain profiles of Au$_{cube}$@Pd$_{cube}$ NPs with a range of Pd shell thicknesses across three Au core sharpness index values (**Fig. 2a-c**). The strain and lattice constant ratio maps represent the average of four edges from a single particle to reduce noise. For sharp-core Au$_{cube}$@Pd$_{cube}$ particles (*SI* = 0.83), the Pd shell maintained a coherent lattice with the Au core for all thickness values explored (**Fig. 2a**), where Au lattice is used as a reference in the $\varepsilon_{xx}$ maps. However, with increasingly rounded cores (*SI* = 0.68, 0.32), strain release in the Pd occurs in thinner shell thicknesses (**Fig. 2b** and **c**). Histograms of the actual shell strain (bulk Pd lattice constant as the reference) in each of these samples further confirms that sharp-tipped cores stabilize strained shells compared to traditional rounded particles (**Fig. 2d**).

Taken together, these data allow for quantification of the *critical thickness* as a function of particle morphology. Here, the thickness specifically refers to the Pd shell thickness on {100} facets. By measuring strain in NPs with different Pd shell thicknesses (1 - 25 nm) for each core geometry (*SI* = 0.83, 0.68 and 0.32), a sigmoidal relationship between Pd strain and shell thickness is observed (**Fig. 2e**). For each geometry, we collected >20 datapoints to ensure it is statistically significant (raw data shown in **Fig. S9-11**). We used the half maximum of the second derivative from these sigmoidal strain plots (defined in **Fig, S12**) to identify the initiation of strain relaxation with high confidence. A threshold value of 4.1% strain in the shell is observed for each system and the Pd shell thickness at which this occurs represents the *critical thickness* for the core-shell $Au_{cube}@Pd_{cube}$ NPs. The measured *critical thickness* for rounded cores (*SI* = 0.32) is 3.1 nm, consistent with previously reported values(*12*, *13*). However, for sharper cores (*SI* = 0.68 and 0.83), the *critical thickness* increases to 6.8 nm and 11.3 nm. The dependence of *critical thickness* on the *SI* values presents a significant increase (as shown **Fig. 2f**), signifying the control of morphology to stabilize epitaxial strain. Importantly, measuring the *critical thickness* for particles with core sizes of 30 nm and 20 nm (**Fig. S13**) shows that this finding tends to be independent of particle size, and is very likely a geometric effect. However, achieving ultra-sharp cores (SI > 0.8) for 30 nm and 20 nm core-size particles poses a large challenge. A tradeoff emerges between particle size and strain stability, with smaller NPs offering more surface area, while larger NPs exhibit a more stable surface strain.

To understand how core geometry facilitates surface strain preservation, we use molecular dynamics (MD) simulations to replicate the layer-by-layer synthesis of core-shell $Au_{cube}@Pd_{cube}$ structures with both sharp and round Au cores (see details for the simulation in **Methods**). The results of the whole-particle MD simulations show that dislocations nucleate around the corners, but at different shell thicknesses for sharp- and round-core cases. In sharp-core particles, the first dislocation nucleates at a 5 nm shell thickness (**Fig. 3a**), whereas rounded particles trigger nucleation with a much thinner Pd shell (<1 nm) (**Fig. 3b**). With an increase in shell thickness, dislocations propagated and multiplied more in round-core particles (refer to **Supplementary Movies**). The calculated dislocation density unequivocally demonstrate a low dislocation density in sharp-core $Au_{cube}@Pd_{cube}$ NPs even beyond 9 nm of Pd shell, while round-core $Au_{cube}@Pd_{cube}$ NPs exhibit a considerably higher dislocation density when the shell grows beyond 1 nm (**Fig. 3c**). The longitudinal surface strain starts to drop in the round-core particles when the dislocation density increases, indicating dislocations facilitate surface strain relaxation (**Fig. 3d**). The results show that misfit dislocations form more easily in round-core nanoparticles, explaining why round-core particles release lattice strain earlier in the synthesis process.

Previous studies have attributed the dislocation formation in face-centered cubic (fcc) metal heterointerfaces to intensive local shear(*28*). In core-shell geometries, large shear stress has been predicted to concentrate at the corners, which causes bulging of atomic planes and later slip to form dislocations in the shell(*12*). Our MD simulations also revealed that the Poisson effect from the Pd shell at {100} facets exert compressive stress on the corner regions (**Fig. 3e** and **f**), leading to localized lattice shear (**Fig. 3g**) and abovementioned lattice bulging (**Fig. 3f**) at the corners. The simulated kinetics also show that dislocations initiate from a void in the surface lattice plane, triggering the formation of a stacking fault bounded by partial dislocations (**Fig. S14**). To understand the effect of corner sharpness on shear stress, strain profiles from sharp- and round-core core-shell NPs with different shell thickness have been simulated (**Fig. S15**). We compared the local shear strain of sharp- and round-core NPs (**Fig. 3i** and **j**), and the results highlight larger shear strain at the corners in the truncated case, which later causes lattice slipping and stacking fault formation. In contrast, the sharp-core case has weaker shear strain, allowing the shell to grow thicker without dislocation nucleation.

To test the stability of surface strain in core-shell Au@Pd nanoparticles, we performed the Suzuki-type homocoupling reactions(*29*) using *trans*-2-phenylvinylboronic acid (PVBA) as a substrate and sharp-core or round-core core-shell $Au_{cube}@Pd_{cube}$ NPs as nanocatalysts. In this class of reactions(*30–32*), Pd undergoes leaching and oxidation into $PdO_x$ species during the catalytic process, followed by re-deposition after formation of new carbon-carbon bonds(*33–37*) (**Fig. 4a**). Previous results indicate this reaction is favored with Pd {100} facets (*38*) under tensile strain(*39, 40*), but the catalytic conditions involve alkaline conditions and an elevated temperature of 80 °C, which can be harsh for catalysts. Consequently, establishing more stable strain states in Pd surfaces would improve the activity and recyclability of catalysts for this heterogeneous reaction(*30, 31, 41*).

To evaluate the strain stability during the catalytic homocoupling, we assessed the strain profile of multiple $Au_{cube}@Pd_{cube}$ NPs before and after the reaction (**Fig. S16**). The results reveal that the sharp-core NPs maintain the epitaxial strain in the shell after the reaction, whereas the strain in round-core NPs is largely released, presumably through the introduction of dislocations (**Fig. 4b**). We evaluated the activity of the core-shell Au@Pd nanocatalysts by monitoring the reaction kinetics via UV-vis spectrophotometry. In this system, the homocoupling products of PVBA are *trans,trans*-1,4-diphenyl-1,3-butadiene (DPBT), which is spectroscopically active with a pronounced absorption peak at $\lambda_{max}$ = 348 nm (calculated in **Fig. S17a**). By monitoring the absorbance of this peak over time, we obtain reaction kinetics for both sharp-core and round-core Au$_{cube}$@Pd$_{cube}$ NPs (**Fig. 4c**). The kinetic profiles show that nanocatalysts with sharp Au cores exhibited a larger slope during the first two hours, indicating a higher catalytic reaction rate in comparison to that of the round-core case. The reactivities of the different Pd shells were determined by comparing the initial reaction rate(*29*) ($\gamma_0$), calculated based on the extinction coefficient of the DPBT (**Fig. S17b**, see **Methods**). Both pure Pd cubes (unstrained, no Au core) and round-core Au$_{cube}$@Pd$_{cube}$ particles above the critical thickness exhibit similar activity due to the relaxed strain in the shell (**Fig. 4d**). In contrast, the sharp-core Au$_{cube}$@Pd$_{cube}$ nanocatalysts demonstrate an activity (6.3 $\times$ 10$^9$ mol Liter$^{-1}$ min$^{-1}$ $m_{Pd}^{-2}$) that is 4.3-fold higher than that for round-core Au$_{cube}$@Pd$_{cube}$ (1.5 $\times$ 10$^9$ mol Liter$^{-1}$ min$^{-1}$ $m_{Pd}^{-2}$) and 5.3-fold higher than pure unstrained Pd cubes (1.2 $\times$ 10$^9$ mol Liter$^{-1}$ min$^{-1}$ $m_{Pd}^{-2}$, **Fig. 4d**). These results confirm that lattice strain stabilized by core-sharpening enhances the reactivity of catalytic surfaces.

Furthermore, we conduct a comparison of the product yields between sharp-core and round-core Au$_{cube}$@Pd$_{cube}$ NP catalysts. Whereas the DPBT yield reaches 74.3% for round-core Au$_{cube}$@Pd$_{cube}$ NPs it is 96.8% with particles synthesized with a sharp core (**Fig. 4e**). In addition, by increasing the thickness of the Pd shell in sharp-core Au$_{cube}$@Pd$_{cube}$ NPs (from 4 nm to 8 nm), the activity and yield remain comparable to the 4-nm sharp-core catalysts (**Fig. 4d** and **e,** green). This observation shows that even with a thicker shell (and greater catalytic surface area per particle), the surface strain of sharp-core core-shell particles remain stable under reaction conditions.

**Discussion**

In summary, we introduce a morphology-control strategy aimed at stabilizing the surface strain of core-shell nanocatalysts. This approach increases the *critical thickness* in core-shell Au$_{cube}$@Pd$_{cube}$

systems when the core morphology is sharper. It accomplishes this by inhibiting dislocation nucleation at the particle corners, which in turn preserve the surface strain. Given that morphology control is more straightforward than consistently producing ultra-thin shells, our work presents a promising method to streamline the synthesis of strain-engineered nanocatalysts with enhanced performance, potentially enabling their large-scale production. This finding holds promise for generalization to other core-shell nanoparticles and even other nano-architectured materials, offering additional opportunities for strain engineered materials.

**Materials and Methods**

**Materials synthesis:**

**Chemical reagents:** Hydrogen tetrachloroaurate hydrate ($HAuCl_4$, 99.999% trace metal basis), *L*-ascorbic acid (AA), sodium borohydride ($NaBH_4$), silver nitrate ($AgNO_3$, 99.999% trace metal basis), palladium chloride ($PdCl_2$, 99.999%), potassium bromide (KBr), Potassium carbonate ($K_2CO_3$), *trans*-2-phenylvinylboronic acid (PVBA), *trans,trans*-1,4-diphenyl-1,3-butadiene (DPBT), hydrogen chloride (37 wt%, 12.1 M), methylcyclohexene (MCH), chloroform are all purchased from Sigma-Aldrich. Hexadecyltrimethylammonium bromide (CTAB) and hexadecyltrimethylammonium chloride (CTAC) were purchased from TCI America. All reagents are used as received. Milli-Q water (0.22 μM pore size, 18.2 MΩ · cm at 25 °C) was used for all syntheses and before each growth, all glassware were treated with aqua regia and rinsed with excess water.

**Synthesis of gold (Au) clusters:** First, $HAuCl_4 \cdot 3H_2O$ solution (10 mM, 250 μL) and 5 mL of 0.2 M CTAB solution are added to 4.75 mL water. Next 0.6 mL of freshly made ice-cold 10 mM $NaBH_4$ solution was quickly injected into the above solution under vigorous stirring. The solution color changed from yellow to brownish, indicating the formation of gold clusters. After vigorously stirring for 2 min and the obtained solution was aged at 27 °C for 3 hours before use.

**Synthesis of 10-nm Au seeds:** Similar to our previously reported works (*42*, *43*), aqueous solutions of CTAC (200 mM, 2 mL), AA (100 mM, 1.5 mL), and an aqueous $HAuCl_4$ solution (0.5 mM, 2 mL) were mixed, followed by one-shot injection of the above Au clusters (50 μL). The reaction was allowed to continue at 27 °C for 15 min. The product was collected by centrifugation at 21000 rcf for 90 min, and then washed with water once for further use and characterization.

**Synthesis of sharp Au cube:** In a typical synthesis of 50 nm gold nanocube with high sharpness, aqueous solution containing CTAC (50 mM, 12 mL), KBr (100 mM, 12 μL), AA (100 mM, 39 μL) was mixed with 10-nm Au seeds (3.406 optical density, OD, 14.5 μL) thoroughly; Afterwards, aqueous $HAuCl_4$ solution (10 mM, 0.3 mL) was added and then kept at room temperature for 3 hours to finish the reaction. It is noteworthy to mention that the size of Au cube NPs can be tuned by the amounts of seeds. By adjusting the ratio between growth solution ($Au^{3+}$) and seeds ($Au^0$) added in the solution, we can achieve different particle sizes. The higher $Au^{3+}/Au^0$ ratio is, the larger particle can be achieved. The size variation of the Au cubes is less than 4% **(Fig. S18)**. In addition, the sharpness can be controlled by changing the concentration of bromide ion in the growth solution, which is independent from the size control. After reaction completion, the obtained Au cube NPs was centrifuged at 6000 rcf for 12 min and washed with water once. The obtained pellet was re-dispersed in 2 mL water for future use.

**Synthesis of truncated Au cube:** In a typical synthesis of 50 nm gold nanocube with truncated tips, two steps are involved. Firstly, the 10-nm Au seeds were firstly grown into larger seeds by introducing seeds (3.406 OD 14.5 μL) into growth solution containing CTAC (200 mM, 1.4 mL), H2O (1.33 mL), AA (100 mM, 1.05 mL) and $HAuCl_4$ (10 mM, 70 μL). After mixing thoroughly, this solution was maintained at room temperature for 30 min and then centrifuged at 10000 rcf for 15 min to get pellets. Afterwards, the pellet was suspended in CTAC (50 mM, 1.5 mL). To which, KBr (100 mM, 3 μL), AA (100 mM, 3.9 μL) and $HAuCl_4$ (10 mM, 30 μL) were added to get truncated cube NPs with blunt tips.

**Preparation of $H_2PdCl_4$ solution:** The $H_2PdCl_4$ (10 mM) was prepared by mixing $PdCl_2$ (44.5 mg) into HCl (0.02 M, 25 mL) aqueous solution under stirring at 50 °C until complete dissolution.

**Synthesis of core-shell $Au_{cube}@Pd_{cube}$ NPs with different Pd thickness:** To epitaxial overgrowth of core-shell $Au_{cube}@Pd_{cube}$ NPs, the Au cube NPs solution (4.86 OD, 500 uL) was added to growth solution containing CTAB (40 mM, 1 mL), $H_2PdCl_4$ (10 mM, 40 μL), AA (100 mM, 9 μL) and then heated at 60 °C for 3 hours to get core-shell $Au_{cube}@Pd_{cube}$ NPs. Similarly, the thickness of Pd shell can be tuned by changing the ratio between growth solution to Au cube NPs and other conditions are the same.

**Preparation Au$_{cube}$@Pd$_{cube}$ catalyst:** In a typical synthesis of sharp Au$_{cube}$@Pd$_{cube}$, sharp Au cube was firstly obtained with above mentioned methods. As a reference, the concentration of NP was determined by the UV-vis spectrum ($\lambda_{max}$ = 556 nm, 4.2 OD, 2 mL). To form Au$_{cube}$@Pd$_{cube}$ NPs, to above Au cube solution, the Pd growth solution containing CTAB (40 mM, 3.6 mL), H$_2$PdCl$_4$ (10 mM, 200 μL) and AA (100 mM, 45 μL) was added, which was heated at 60 °C for 3 hours to get core-shell Au$_{cube}$@Pd$_{cube}$ NPs. After completing the growth, it was centrifuged at 5000 rcf for 5 min to get pellets and the particles were washed with water once to remove any remaining reagent that may affect the catalytic test. Afterwards, the obtained solution was carefully transferred into 1.5 mL centrifuge tube and wash with relatively low centrifugation speed (4000 rcf, 3 min), the supernatant was carefully removed and then all the particles were re-suspended in 200 μL water for future use. To control the thickness of the Pd shell, a different amount of Pd growth solution was added to the Au cube solution. Similarly, for the truncated Au$_{cube}$@Pd$_{cube}$ NPs, a similar growth procedure was used to control the Pd thickness expecting replacing sharp Au cube with the truncated Au cube.

**4D-STEM strain mapping:**

**EMPAD data acquisition:** The 4D-STEM datasets were taken on an aberration-corrected Thermo Fisher Titan Themis at 300 keV with an Electron Microscope Pixel Array Detector (EMPAD). A 1.76-mrad convergence angle was used, leading to a 0.69 nm probe size (defined by Full Width Half Maximum (FWHM) probe diameter). The camera length is 185 mm to capture more higher order diffraction spots for the exit-wave power-cepstrum (EWPC) method. For a 300 kV electron beam, 579 ADUs represent one electron per pixel. For all the datasets, an exposure time of 1.86 ms per frame (1 ms acquisition time along with 0.86 ms readout time) was employed when acquiring the EMPAD 4D datasets. The scan size in real space (the number of pixels the beam scans across) can be set from 64×64 to 512×512. The scan size of the data used in this work was 256×256. The total time for capturing one 4D data was 122 seconds.

**Strain Mapping:** The EWPC method is used to measure the strain profiles of the core-shell nanocubes. The workflow of EWPC method is shown in **Fig. S2a**. The raw diffraction pattern (**Fig.S2b**) is first transformed to logarithmic scale (**Fig. S2c**) and then a Gaussian mask (**Fig. S2d**) is applied on it to reduce the boundary effects (**Fig. S2e**). The power cepstrum image is acquired through fast-Fourier transform (FFT) of the diffraction (**Fig. S2f**). The cepstrum spot positions are measured through the center of mass method to determine the two lattice vectors along x and y directions. The strain profiles, including $e_{xx}$, $e_{yy}$, $e_{xy}$ and rotation can be calculated through polar decomposition method. The strain maps in **Fig. 2a-c** average edges with similar thickness (variance < 1 nm) to improve the signal to noise ratio.

The averaged lattice constant of center Au core region is used as the reference of the strain map. When the Pd shell is thin and strained, the center region reference is regarded as the lattice constant of bulk Au (4.078 Å). When the Pd shell is thick and the strain is released, the center region is regarded as a mixture of bulk Au and Pd lattice overlayed, owing to the projection effects from the transmitted electron beam. Therefore, we perform the strain correction method shown in **Fig. S3**. The center diffraction and cepstrum is expressed as the Au + x * Pd (**Fig. S3a**), where the coefficient x is decided by the thickness and strain of the Pd shell. We calculate the diffraction difference between cepstrum of the center Au region and Pd shell (**Fig. S3b**), where a dipole shows at each cepstrum spot (**Fig. S3c**). The blue and red contrast in the dipole represent the Au and Pd lattice, respectively (**Fig. S3d**). Through tuning the coefficient (x) of the cepstrum from the Pd shell, the Pd peak intensity in the dipoles at cepstrum spots reduces (**Fig. S3e**). The coefficient (0.11), which gives the zero Pd peak intensity, are chosen to correct the cepstrum in the Au region for a more precise strain mapping. After the correction, the lattice constant in the center region of the thick Pd shell particle can also be regarded as the bulk Au lattice constant.

To determine the *critical thickness*, the shell strain and shell thickness are measured statistically from a number of core-shell NPs with different thickness and core sharpness. All measured data points are displayed in **Fig. S9**. The shell strain-thickness scattered plots are fitted by sigmoid functions, which is a commonly used "S shape" function. The expression is the following:

$$\epsilon = 4.8\% \times \frac{1}{1 + e^{0.5(t-c)}} + 4.8\%$$

Where $\epsilon$ is the shell strain and the t is shell thickness. 4.8% is the difference between the unstrained Au and Pd lattice. By performing the least square fitting of the scattered data points to the sigmoid function, the parameter c, which represent the center of the sigmoid function, can be determined for each core-sharpness NPs. The derivations of the fitted curves are calculated and shown as the colored shadows in **Fig. S12**. The thickness at the half of the minimum derivation points is chosen as the critical thickness where the strain starts to release (**Fig. S12**).

**Molecular Dynamics Simulation:** Seed atomistic models are generated from an infinite fcc lattice with cell parameter 4.08Å by selecting all lattice sites that fall within the outline of the NP. These sites are occupied with Au and Pd atoms according to the core-shell nanostructure geometry. Atoms in the Pd shell are sorted into concentric monolayers (**Fig. S19a**). Layers are successively added to the MD model in a series of simulation stages, reproducing the kinetics of Pd shell growth onto the Au core (**Fig. S19b-c**). The model is simulated with classical MD using the LAMMPS software package(*44*). Atom velocities are sampled from the Maxwell-Boltzmann distribution at temperature 300 K. Interatomic potentials are computed with the long-range generalized form of the embedded-atom method (EAM) of Finnis and Sinclair using the quantum Sutton-Chen force field(*45–49*). During each deposition stage, the MD models are relaxed for 0.1 ns at 300 K using a Langevin thermostat with a dumping constant of 0.0020 fs and a 1 fs time Integration in the microcanonical ensemble.(*50*)

After energy minimization of the Au core, Pd layers are deposited layer-by-layer onto the surface of the MD models while keeping the atom dynamics in equilibration. The procedure to grow the Pd shell by one layer is as follows: First, a new Pd layer is created surrounding the existing NP. To mimic the effect of the bromide concentration in experiment, we temporarily include an auxiliary Pd layer on the outside of the new Pd layer (**Fig. S19b**). This auxiliary Pd layer ensures stability of the NP surface by suppressing diffusion of Pd atoms during deposition. Its presence reproduces the even and smooth surface growth observed in experiment as a consequence of the presence of bromide. Second, all Pd atoms of the new layer are moved radially from their initial positions towards the NP center up to a minimum distance of 3 Å from any previously equilibrated atom. Atoms in the auxiliary layer are also moved radially up to a minimum distance of 2 Å from the atoms in the new layer. The smaller cut-off distance promotes inwards dynamics of the atoms improving deposition kinetics. Third, the system is equilibrated using MD simulation. Fourth, the atoms of the auxiliary layer are removed from the simulation (**Fig. S19c**).

Snapshots of the MD simulation are recorded at 10 ps time intervals. The symmetric and antisymmetric strain components are computed from the strain tensor measured with the cell deformation method (*51*, *52*). The OVITO software package is used to render the MD snapshots, to identify the local structure coordination, and to measure the dislocation density (*53–57*).

**Catalytic reaction:** To test the catalytic activities of both sharp Au cube@Pd cube and round Au cube@Pd cube, the *trans*-2-phenylvinylboronic acid (PVBA) homocoupling reaction is performed in both cases. To determine the reaction kinetics of this reaction, *trans*-2-phenylvinylboronic acid (PVBA), K2CO3, water and methylcyclohexene (MCH) were firstly transferred into quartz cuvette with an optical path of 10 mm. This cuvette is loaded into a preheated UV-vis stage (80 °C) for 10 min before adding Au@Pd catalyst (50 uL), which is prepared according to the previous section. All the catalytic measurements were conducted with the condition as follows: [PVBA] = 0.40 mmol L$^{-1}$, and [K$_2$CO$_3$] = 2.00 mmol L$^{-1}$, in water/MCH (1/1) biphasic medium. The reaction kinetics of the PVBA homocoupling were investigated by monitoring the appearance of the product (DPBT) at wavelength of 348 nm with UV–vis spectrophotometry. The reactivity of the catalyst is determined by the initial reaction rate $\gamma_0$, which is determined by the first two points in the plot and the normalized by the total surface area of Pd cube; the yield is determined by normalizing the equilibrium adsorption intensity with the theoretical adsorption of the DPBT.

**Acknowledgements**: C.S., Z.C., and Y.H. acknowledge the use of Electron Microscopy Center (EMC) at Rice University. A.L. acknowledges Lilly Endowment, Inc. through its support for the Indiana University Pervasive Technology Institute. **Funding**: The work is supported by NSF (CMMI-2239545). C.S. and Y.H. acknowledge the support from the Robert A. Welch Foundation (C-2065), and American Chemical Society Petroleum Research Fund (67236-DNI10). M.R.J. thanks the Robert A. Welch Foundation (C-2146), the David and Lucile Packard Foundation (2018-68049), and the American Chemical Society Petroleum Research Fund (65837-ND3). A.L. and M.E. acknowledge support from the Deutsche Forschungsgemeinschaft (DFG, German Research Foundation) Project-ID 452477982–LE4543/2-1. **Author contributions**: Conceptualization: M.R.J, and Y.H. Data curation: C.S., Z.C., A.L. Methodology: C.S., Z.C., A.L. Investigation: C.S., Z.C., A.L. Visualization: C.S., Z.C., and A.L. Funding acquisition: M.E, M.R.J, and Y.H. Project administration: M.R.J, and Y.H. Supervision: M.E, M.R.J, and Y.H. Writing—original draft: C.S., Z.C., Y.H. Writing—review and editing: C.S., Z.C., A.L., Y.Y., M.E, M.R.J, and Y.H. **Competing Interests**: All authors declare no competing interests. **Data and Materials Availability:** All data are available in the main text or the supplementary materials.

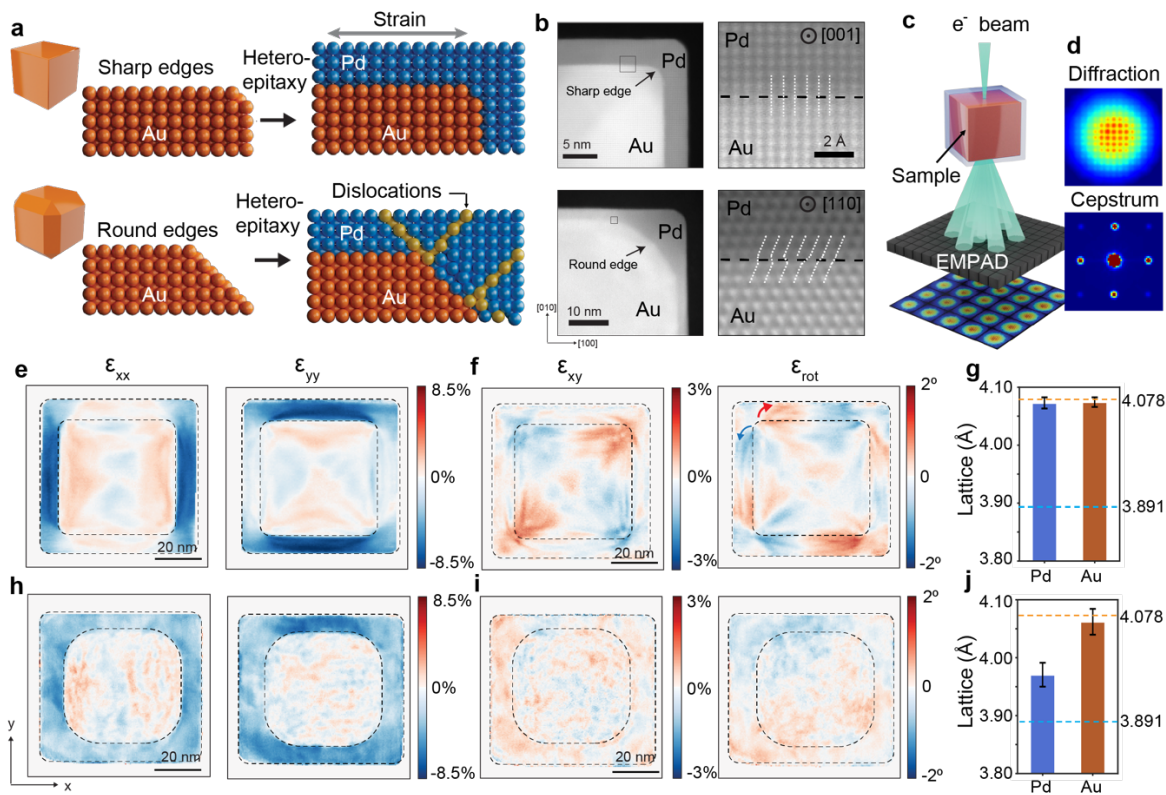

**Figure 1. Enhancing surface strain stability *via* nanoscale morphology control.** (a) Schematic illustration of strain preservation by a nanoscale morphology with sharp edges. (b) Atomic-resolution ADF-STEM images of the sharp-core (top left) and round-core (bottom left) Au$_{cube}$@Pd$_{cube}$ NPs. Zoomed-in images from boxed area show a coherent interface (top right) and lattice mismatch (bottom right) between Au and Pd in sharp-core and round-core particles respectively, with the zone axes labeled. (c,d) Schematic showing 4D-STEM with a diffraction pattern (d, top) and an EWPC (d, bottom) from the core-shell NP for precise strain analysis. (e,f,h,l) Strain maps ($\varepsilon_{xx}$, $\varepsilon_{yy}$, $\varepsilon_{xy}$ and $\varepsilon_{rot}$) from individual sharp-core (e,f) and round-core (h,l) Au$_{cube}$@Pd$_{cube}$ NPs. (g,j) Lattice parameters that is parallel to the interface between Au and Pd for sharp-core (g) and round-core (j) NPs.

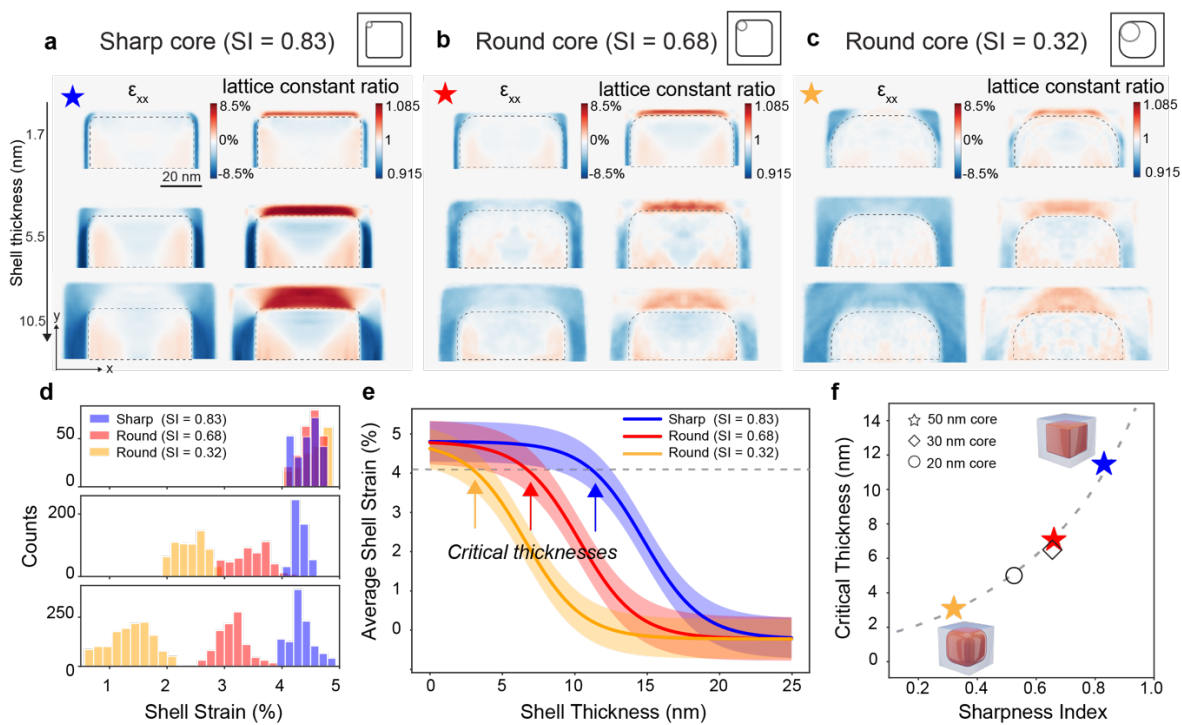

**Figure 2. Morphology-dependent *critical thicknesses*.** (a-c) 4D-STEM strain maps ($\varepsilon_{xx}$ with Au lattice as reference) and lattice-constant-ratio maps of the core-shell $Au_{cube}@Pd_{cube}$ particles with different shell thicknesses (top to bottom 1.7±0.3 nm, 5.5±0.5 nm, and 10.5±0.5 nm, labeled aside) and different core sharpness with *SI* = 0.83 (a), 0.68 (b), and 0.32 (c). (d) Histogram of strain measurements in Pd shells from the samples in (a-c) with thickness increasing top to bottom. Bulk Pd lattice is used as the reference. (e) Average shell strain with increasing shell thickness for three core geometries. (f) *Critical thickness* of core-shell $Au_{cube}@Pd_{cube}$ NPs depending on the core sharpness and particle size.

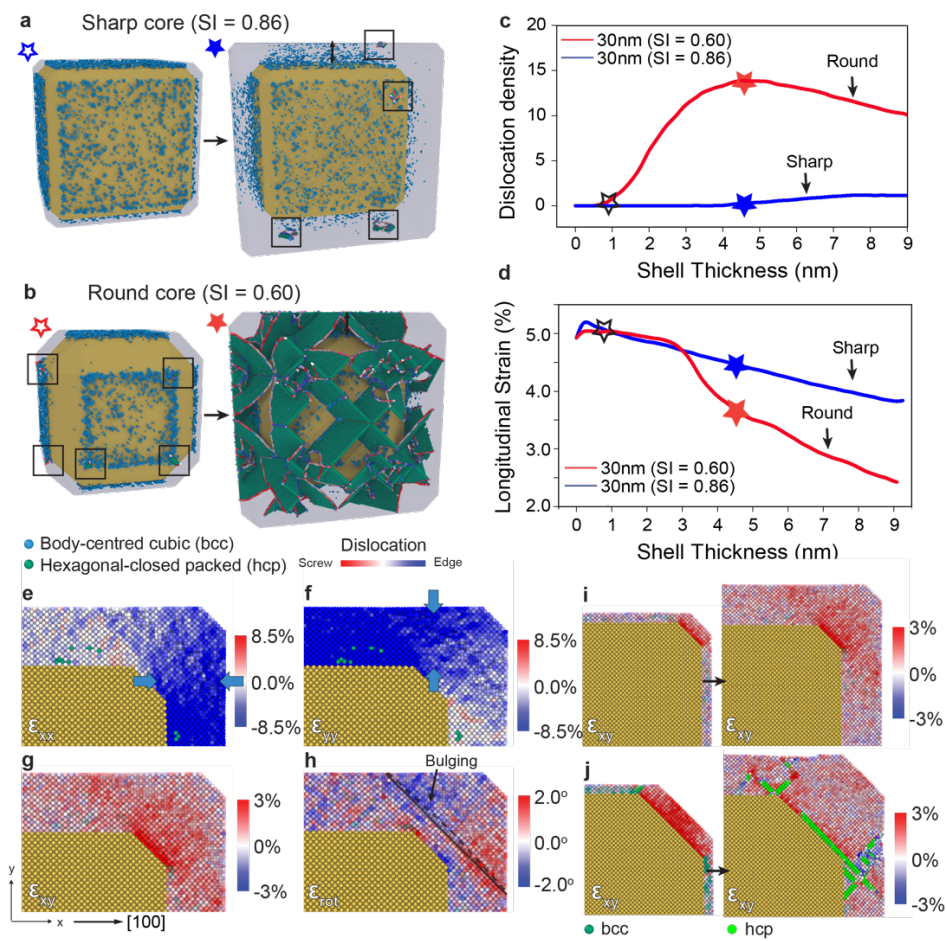

**Figure 3. Molecular dynamics simulation of dislocation nucleation. (a,b)** 3D views of core-shell Au$_{cube}$@Pd$_{cube}$ MD simulations with different shell thicknesses (1.0 nm and 5.0 nm) and different core sharpnesses (SI = 0.86 (a) and 0.60 (b)). Because the Bain transformation path indicates that deformation of the face-centered cubic (fcc) structure yields a centered cubic coordination, we identify atoms with hcp and body-centered cubic (bcc) local order in different colors, where bcc highlights regions with large fcc structure deformation. **(c)** Dislocation density (total length over volume) as a function of the shell thickness from MD simulations. **(d)** Average longitudinal strain ($\varepsilon_{xx}$) at the center of the particles' facets in a radius of 5 nm. **(e-h)** Strain maps of a slice through a simulation snapshot for a Au$_{cube}$@Pd$_{cube}$ NP (SI = 0.86) with ~4 nm shell thickness. In (h), the dotted line traces the atomic column, which deviates from a straight line (solid), indicating the lattice bulging. **(i,j)** Shear maps from simulation snapshots from sharp (SI = 0.86) (i) and truncated (SI = 0.60) (j) NPs at a thin (~1 nm) and thick (~4 nm) shell thickness. Shear and rotation maps are averaged over five sub-slices to reduce noise.

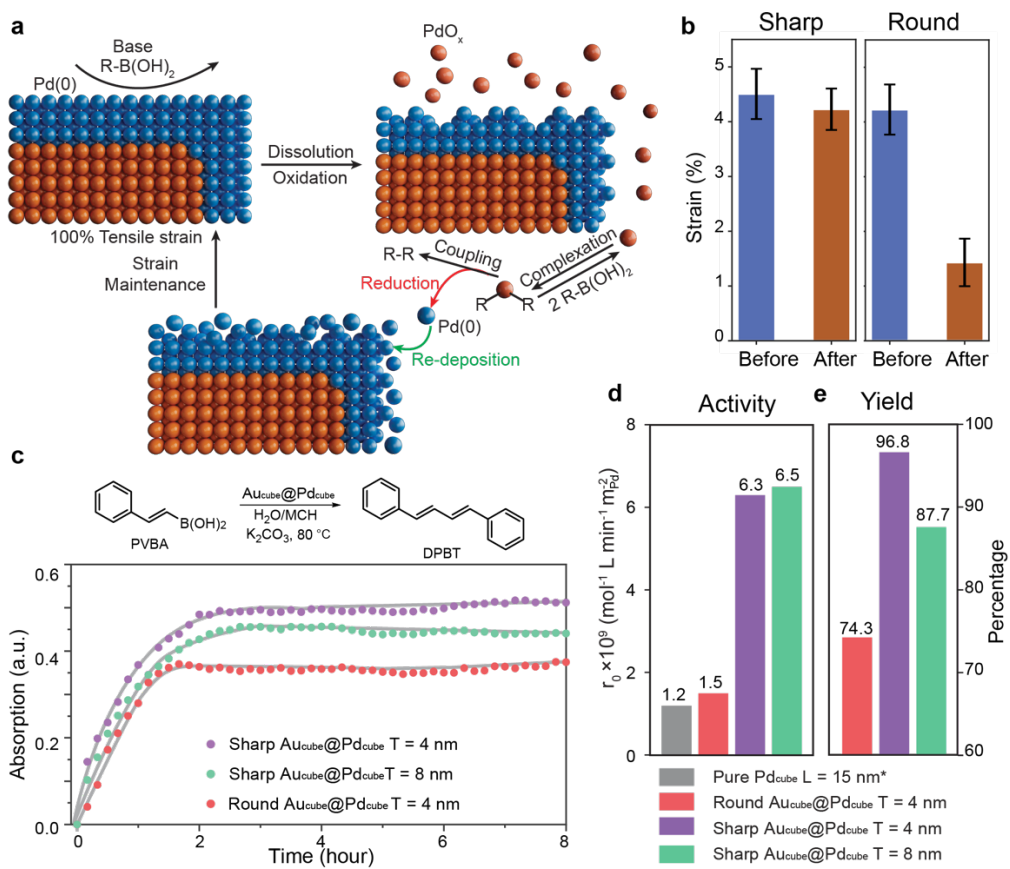

**Figure 4. Strain stability under catalytic reactions. (a)** Schematic illustration of the catalytic process of homocoupling in a Suzuki-type reaction on sharp-core core-shell $Au_{cube}@Pd_{cube}$ NPs with tensile strain. This type of reaction relies on the dissolution, reduction, and re-deposition of Pd, which presents a harsh environment for strain engineered catalysts. **(b)** The average strain in sharp- and round-core $Au_{cube}@Pd_{cube}$ NPs with a 4 nm shell thickness before and after the catalytic reaction. **(c)** The kinetic profiles of homocoupling reaction from $Au_{cube}@Pd_{cube}$ catalysts with sharp (purple and green) and round cores (red). Top: homocoupling reaction catalyzed by Pd surfaces. **(d)** Comparison of the catalytic activities of the pure $Pd_{cube}$ (gray)[27] with round-core (red) and sharp-core (purple and green) $Au_{cube}@Pd_{cube}$ nanocatalysts. The rate is normalized by the surface area of Pd. **(e)** Comparison of the yield between round-core (red) and sharp-core (purple and green) $Au_{cube}@Pd_{cube}$ nanocatalysts.